\documentclass{ieeeaccess}
\usepackage{cite}
\usepackage{amsmath,amssymb,amsfonts}
\usepackage{algorithmic}
\usepackage{graphicx}
\usepackage{textcomp}
\def\BibTeX{{\rm B\kern-.05em{\sc i\kern-.025em b}\kern-.08em
    T\kern-.1667em\lower.7ex\hbox{E}\kern-.125emX}}

\usepackage{xspace}
\usepackage{lmodern}

\newcommand\blfootnote[1]{%
  \begingroup
  \renewcommand\thefootnote{}\footnote{#1}%
  \addtocounter{footnote}{-1}%
  \endgroup
}

\newcommand{\quot}[1]{``#1''}

\newcommand{\qnick}[0]{QuanvNN\xspace}
\newcommand{\qnicks}[0]{{\qnick}s\xspace}

\newcommand{\lartpc}[0]{LArTPC\xspace}
\newcommand{\mirabest}[0]{MiraBest\xspace}

\newcommand{\qnnrotational}[0]{QNN-\textsc{Rot}\xspace}
\newcommand{\qnnintegrated}[0]{QNN-\textsc{Int}\xspace}

\newcommand{\RNDLIN}[0]{\textsc{RndLin}\xspace}
\newcommand{\RNDMUL}[0]{\textsc{RndMul}\xspace}
\newcommand{\SIMPLE}[0]{\textsc{Simple}\xspace}

\newcommand{\qnnintegratedRNDLIN}[0]{QNN-\textsc{Int-}\RNDLIN}
\newcommand{\qnnintegratedRNDMUL}[0]{QNN-\textsc{Int-}\RNDMUL}
\newcommand{\qnnintegratedSIMPLE}[0]{QNN-\textsc{Int-}\SIMPLE}

\newcommand{\encrotational}[0]{rotational encoding\xspace}

\newcommand{\articletitle}[0]{Integrated Encoding and Quantization to Enhance Quanvolutional Neural Networks$^*$}

\begin{document}
\history{Date of publication xxxx 00, 0000, date of current version xxxx 00, 0000.}
\doi{10.1109/TQE.2020.DOI}

\title{\articletitle}

\author{%
\uppercase{Daniele Lizzio Bosco}\authorrefmark{1,2},
\uppercase{Beatrice Portelli\authorrefmark{1,2},
and
Giuseppe Serra}.\authorrefmark{1}%
}

\address[1]{Department of Mathematics, Computer Science and Physics, University of Udine, UD 33100 Italy (email: lizziobosco.daniele@spes.uniud.it, portelli.beatrice@spes.uniud.it, giuseppe.serra@uniud.it)}
\address[2]{Department of Biology, University of Naples Federico II, NA 80126 Italy}


\markboth
{Lizzio Bosco \headeretal: \articletitle}
{Lizzio Bosco \headeretal: \articletitle}

\corresp{Corresponding author: Daniele Lizzio Bosco (email: lizziobosco.daniele@spes.uniud.it).}

\begin{abstract}
Image processing is one of the most promising applications for quantum machine learning (QML). Quanvolutional Neural Networks with non-trainable parameters are the preferred solution to run on current and near future quantum devices.
The typical input preprocessing pipeline for quanvolutional layers comprises of four steps: optional input binary quantization, encoding classical data into quantum states, processing the data to obtain the final quantum states, decoding quantum states back to classical outputs.
In this paper we propose two ways to enhance the efficiency of quanvolutional models.
First, we propose a flexible data quantization approach with memoization, applicable to any encoding method. This allows us to increase the number of quantization levels to retain more information or lower them to reduce the amount of circuit executions.
Second, we introduce a new integrated encoding strategy, which combines the encoding and processing steps in a single circuit. This method allows great flexibility
on several architectural parameters (e.g., number of qubits, filter
size, and circuit depth) making them adjustable to quantum
hardware requirements. 
We compare our proposed integrated model with a classical
convolutional neural network and the well-known rotational encoding method, on two different classification tasks.
The results demonstrate that our proposed model encoding exhibits a comparable or superior performance to the other models while requiring fewer quantum resources.
\end{abstract}


\titlepgskip=-15pt

\maketitle

\section{Introduction}
\blfootnote{* This work has been submitted to the IEEE for possible publication. Copyright may be transferred without notice, after which this version may no longer be accessible.}
The field of Quantum Machine Learning (QML) applied to Computer Vision has gathered increasing interest in the last decade, combining quantum computing and machine learning {to develop new algorithms which may lead to more efficient and optimized computer vision models} \cite{QML_4image_processing_survey}.
A promising approach in image processing is the application of Quantum Convolutional Neural Networks, also known as \textit{Quanvolutional} Neural Networks (hereafter \qnick), which aim to enhance classical models through hybrid (classical-quantum) architectures.
However, current quantum devices are still characterized by a limited number of qubits and {the absence of} error correction. This hinders QML from matching the performance of classical ML methods.
Therefore, the applications of \qnicks are currently limited to simple architectures and small datasets because of several constraints, including: the low number of available qubits, the need to reduce circuit depth to avoid decoherence, and technical optimization constraints. For instance, while deep learning relies on gradient descent for parameter updates, quantum neural networks require the use of the parameter shift rule \cite{Gradient_2019zff}, which involves a large amount of additional circuit measurements and is unreliable in absence of error correction. This makes large scale \qnick optimization currently impractical.

To avoid the need to optimize quantum circuit parameters, a solution is the use of quanvolutional layers with \textit{non-trainable parameters}, as introduced in \cite{Henderson2019QuanvolutionalNN}.
These layers can be used in hybrid models for preprocessing the input dataset, acting as random feature extractors which might identify features that are challenging to compute using classical methods.
Some examples of successful applications, where quantum methods employing trainable quantum parameters would be impractical on current quantum or simulated devices, include speech recognition \cite{speech}, building damage assessment \cite{bhatta_multiclass_2024}, medical diagnostics \cite{arrhythmia}, and pollution emissions monitoring \cite{atmos14060944}. 


The key components of \qnicks are the \textit{quanvolutional layers}, the quantum equivalent of the classical convolutional layers, which process the input in a locally-invariant fashion, detecting patterns and extracting meaningful information.
Typically, quanvolutional layers with non-trainable parameters are the first layers of a \qnick and their input preprocessing comprises of the following stages (see Fig. \ref{fig:full_pipeline}):

\begin{itemize}
\item Binary \textit{Quantization}: if needed, the original input image is converted to a new image with binary pixel values.
\item \textit{Encoding}: the classical input is transformed into quantum states.
\item \textit{Processing}: the quantum states are processed through a pre-defined quantum circuit to create interaction between the input features.
\item \textit{Decoding}: classical information is extracted from the final quantum state of the processing circuit.
\end{itemize}

\Figure[htbp](topskip=0pt, botskip=0pt, midskip=0pt)[width=\linewidth,  trim={0 1cm 7cm 0}, clip]{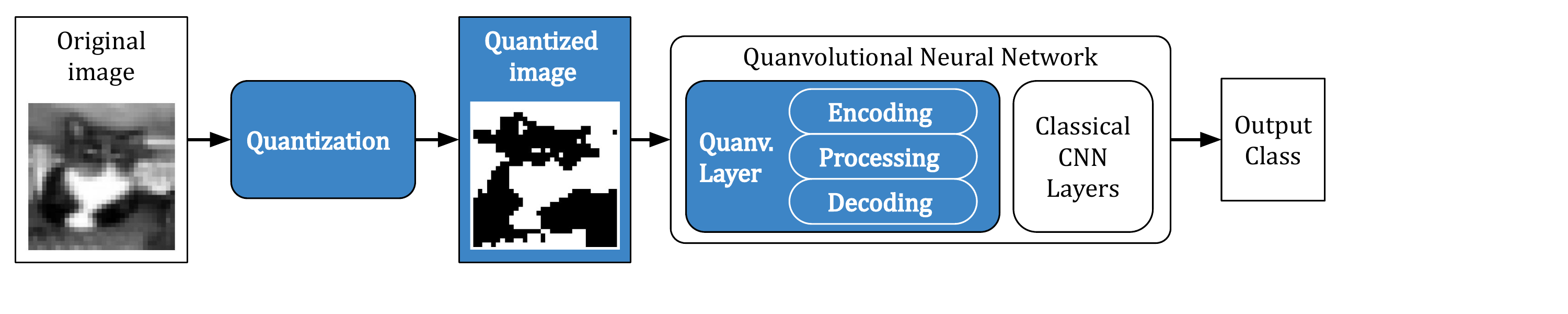}
{Schema of the full pipeline using a quanvolutional neural network. The input image is first quantized and then passed to the network, which comprises a quanvolutional layer (encoding, processing, and decoding) and one or more classical convolutional layers which provide the final output. \label{fig:full_pipeline}}

Despite them being more NISQ-friendly than \qnicks with learnable parameters, quanvolutional layers still present some challenges that need to be addressed.

For example, binary quantization simplifies the input data to decrease the number of required quantum circuit executions, however it may lead to considerable information loss for some tasks which rely on fine-grained information. 
While some encodings do no require quantization and work on the original image, several strategies need binary quantization as a pre-requisite, such as threshold encoding \cite{Henderson2019QuanvolutionalNN}.



On the other hand, encoding and processing stages are the main source of the quantum hardware requirements for quanvolutional neural networks.
For example, rotational encoding \cite{schuld} requires a number of qubits which is quadratic with respect to the size of the input patch, usually denoted as \textit{kernel size}. As an example, a kernel of size $5$ requires $25$ qubits.
Other encodings are even more resource-expensive, such as, amplitude encoding \cite{zheng_design_2023, wang_development_2022} which requires an exponential number of gates to prepare the quantum states. These techniques are unsuitable for Noisy Intermediate-Scale Quantum (NISQ) devices, both for the required number of input qubits and their circuit depth which may cause problems due to limited coherence time.

\medskip

To address these challenges, we propose a new quanvolutional layer which makes the quantization, encoding, and processing stages more resource-efficient and NISQ-friendly while retaining the performance of more expensive quanvolutional models.

As regards \textit{quantization} step, we implement a flexible quantization approach coupled with a look-up table.
The number of quantization levels can be analytically chosen based on the amount of information loss on the input data.
This allows increasing (or decreasing) the number of quantization levels as needed depending on the application, deciding between retaining more information (more quantization levels) or lowering the number of circuit executions and measurements (less quantization levels). 

With respect to the \textit{encoding} and \textit{processing} stages, we propose a quanvolutional model that, differently from previous ones, integrates \textit{both} steps in a single circuit. The proposed \textit{integrated encoding} has no constraints on the number of qubits and gates needed to process the input, and can therefore be adapted to the available resources. This encoding allows creating small and efficient models which are NISQ-friendly, but also scaling up the number of qubits and gates as new resources become available.

We compare the proposed integrated model with a quanvolutional model using \encrotational introduced in \cite{Henderson2019QuanvolutionalNN} and used in \cite{speech, bhatta_multiclass_2024, arrhythmia, atmos14060944, Quanv_radiology, QML_image_by_Senokosov}, as well as a classical Convolutional Neural Network (CNN) model. The performance of all models is tested on two different datasets for binary (\mirabest \cite{mirabest_article}) and multi-class image classification (\lartpc \cite{ChenHEPQCN}), using different circuit configurations and testing different kernel sizes for the input patches.

Our experiments show that the proposed integrated encoding surpasses the performance of the traditional \encrotational for all kernel sizes. Additionally, integrated encoding is more resource-efficient and could therefore be tested on larger kernel sizes compared to \encrotational. Finally, our proposed approach outperforms the classical CNN in both tasks, while the model using \encrotational only does so on the binary classification benchmark.

In addition, our experiments show that, contrary to standard quanvolutional models, the expressivity of our proposed integrated model changes with the number of gates used. This descriptor is correlated with increased performance on QNNs for classification tasks \cite{exprCorrelation}, but also with certain challenges pertaining to model trainability, such as \textit{barren plateaus} \cite{connecting_expr_and_barren_plateaus}.

To encourage further research on this topic, we share all the code to reproduce our experiments\footnote{\underline{https://github.com/Dan-LB/integrated\_encoding\_for\_}\break\underline{QuanvNN}}, including the implementation of the quantization procedure and the quanvolutional model with integrated encoding.

\medskip

The following sections are organized as follows:
in Section 2, we present a review of the existing literature on the topic.
In Section 3, we provide a concise background of the technical concepts used in this work, including quanvolutional layers and standard encoding strategies.
In Section 4, we introduce our proposed pipeline, describing the data quantization method and the proposed integrated encoding.
Section 5 contains the details of the experimental setting, including datasets, tested models, and implementation details.
Section 6, presents and discusses the experimental results, while Section 7 contains the closing remarks.

\section{Related Work}

Previous research has developed two main classes on quanvolutional methods: the ones where quantum circuits parameters have learnable parameters, and the methods where they have fixed non-learnable parameters (i.e. only the classical parameters are learned).

\subsection{\qnicks with learnable parameters}


These quantum convolutional models include parameterized circuits, which can be trained together with the classical parameters via the parameter shift rule or similar techniques.

The most common approach consists in encoding the input image in a quantum state, processing it via one or more quantum convolutional layers, decoding the final state, and finally passing it to a classical architecture.

For example, \cite{ChenHEPQCN} implemented this kind of \qnick for the classification of High Energy Physic particles on the \lartpc dataset.
Their model encodes each pixel of a $2\times2$ patch in a qubit by converting it in rotation angles. The processing circuit has 4 qubits and a fixed architecture which contains parameterized rotations. These parameters are initialized randomly and iteratively optimized during training. The output of the quantum circuit is then decoded and passed to a series of classical fully connected layers, which are jointly trained with the quantum circuit.

Other works apply the same architecture to the classification of 2D radiological images \cite{Quanv_radiology} and several multi-class classification tasks on MNIST, Medical MNIST, and CIFAR-10 \cite{QML_image_by_Senokosov}. The former tests the effectiveness of several encoding methods which require one qubit per pixel (threshold encoding, rotational encoding, and higher order encoding), while the latter uses rotational encoding.

Another interesting approach consists in placing the quantum circuit between two classical layers. For example, \cite{Quanv_radiology} also introduced a model for the classification of 3D radiological images, which takes $2\times2\times2$ patches as input. The model consists of two classical convolutional layers, followed a quantum convolutional layer with 8 filters. Each filter uses 8 qubits, angle encoding, and contains gates with trainable rotations. Finally, the output of the circuit is processed by further classical fully connected layers as seen above.

 Similar architectures are employed in references \cite{QML_image_by_Senokosov} and \cite{remoteSensing}, where the quantum circuit is preceded by a classical \quot{embedding} phase. This approach has the advantage of decoupling the dimension of the input patches from the dimensions of the circuit: since the input image is first processed by classical layers, these can be used to create a low-dimension embedding which requires less quantum resources to encode and process. On the other hand, this process does not allow the quantum circuit to access the real input data. This may hamper the ability of the quantum circuit to detect meaningful non-classical patterns.

\subsection{\qnicks with non-learnable parameters}

Although parameterized circuits are extensively used in the literature, the process of optimizing their parameters is currently very resource-expensive. For this reason, several works focused on \qnicks with non-learnable parameters. These parameters are usually randomly initialized and remain fixed, while the classical layers of the network undergo the standard training procedure. Another advantage of used a non-learnable quanvolutional layer is that its output can be pre-computed for all the input patches in the dataset, further reducing the amount of quantum circuit executions required to train the model.

This kind of \qnicks was first introduced by Henderson in \cite{Henderson2019QuanvolutionalNN}. In this work, the authors use a classical CNN preceded by a quanvolutional layer. The authors use threshold encoding on $3\times3$ input patches, meaning that the processing circuit requires 9 qubits. The processing circuit is composed by a sequence of randomly-selected parametric and non-parametric gates (more details in Section \ref{sec:henderson_processing}). Any required gate parameter is also selected randomly. The authors test networks consisting of 1 to 50 quanvolutional filters, showing that the performance of the \qnick increases with the number of filters, but this effect caps at around 25 filters.

Following works used similar architectures for the assessment of seismic damage from photos \cite{bhatta_multiclass_2024} and the detection of arrhythmia from ECG signals \cite{arrhythmia}. Both works employed rotational encoding instead of threshold encoding, allowing to encode a wider range of information. They also used smaller $2\times2$ input patches, resulting in a quantum circuit with 4 qubits, followed by classical convolutional and fully connected layers.

Finally, the authors of \cite{quanv_better_enc} compare \qnicks models with randomly-generated circuits with both learnable and non-learnable parameters.
The authors compare the models created from all possible combinations of three different encodings (FRQI, NEQR, and threshold encoding), two input patch sizes ($2\times2$ with 2 filters and $4\times4$ with 4 filters, leading to a qubit requirement of 3 and 16 respectively), and two settings for the parameters (learnable and non-learnable).
The final results show that both learnable and non-learnable circuits can achieve high performances on the MNIST dataset (over 0.82 accuracy). In addition, larger input patch sizes do not necessarily lead to higher performances, as all tested models reached higher accuracy when using $2\times2$ patches.

The \qnick models used in this work are also based on randomly-generated non-learnable quantum circuits. Differently to previous works, we further test the effect of using different input patch sizes (from $2\times2$ to $5\times5$). In addition, we compare the performance of the traditional combination of rotational encoding and Henderson-based processing circuit with the proposed integrated encoding, which shows increased performance with less quantum requirements.

\section{Background}

\subsection{Quanvolutional Layers}

Quanvolutional layers are the fundamental component of \qnicks, and are based on classical convolutional filters, which have transformed the field of image processing and computer vision. Similarly to their classical counterparts, they extract meaningful features from images in a locally space-invariant manner, but they also exploit the capability of quantum circuits to extract complex features that are difficult to obtain classically.

Quanvolutional layers comprise of one or multiple quanvolutional filters, which perform operations on a local subsection of the input data through quantum circuits.

A filter, also called \textit{kernel}, maps a subsection of $k \times k$ input data (pixels) $x_1,\dots,x_{k^2}=\mathbf{x}$ to a single scalar value. The input $\mathbf{x}$ is usually referred to as a \quot{patch}. In the classical approach, this mapping is performed using classical operations, such as a scalar product between the patch values and the filter's weights, and the addition of a bias. A quantum convolutional filter acts in a similar manner, with the important difference that the mapping is performed through a quantum circuit. 

More in detail, as mentioned in the Introduction, each filter performs the following operations:
\begin{itemize}
    \item \textit{Encoding}: the classical data (i.e. pixels in the patch $\mathbf{x}$) are encoded in a quantum state;
    \item \textit{Processing}: the quantum state representing the classical data is processed through a sequence of gates; 
    \item \textit{Decoding}: classical information is extracted from the final quantum state.
\end{itemize}

The following sections describe the three phases in more detail.

\subsection{Encoding}

In general, encoding strategies are implemented as quantum circuits which are applied \emph{before} the processing step. Their aim is to embed the classical inputs into a quantum state before further processing.

\subsubsection{Rotational Encoding}

The most common approach to data encoding in quanvolutional models is through the use of \textit{\encrotational} with $R_X$ gates \cite{schuld}. The parametric gate $R_X(\theta)$ is defined as \[
R_X(\theta) = \begin{pmatrix}
\cos\left(\frac{\theta}{2}\right) & -i\sin\left(\frac{\theta}{2}\right) \\
-i\sin\left(\frac{\theta}{2}\right) & \cos\left(\frac{\theta}{2}\right)
\end{pmatrix},
\]
and is obtained by the matrix exponential $\exp(-i\theta X)$, where $X$ denotes one of the Pauli gates. This approach is also known as angle encoding.

In a quanvolutional filter of size $k\times k$, each pixel $p_i$ is encoded on a different qubit. This implies that encoding $k^2$ pixels requires a circuit with exactly $k^2$ qubits.
Consequently, as $k$ increases, this approach becomes unsustainable for devices with a lower qubit count. Conversely, it may also reduce the expressivity of the filter for lower values of $k$.  

As the matrix exponential has a period of $2\pi$, each pixel is mapped to a rotation angle before the encoding. For example, if $p_i \in [0, 1]$, then $p_i$ is encoded as $R_X(p_i\pi)$ applied to the $i$-th qubit. Formally, the mapping of a patch of size $k\times k$ to the corresponding quantum state can be written as 

{
\begin{align}
    &|0\rangle^{\otimes n} \mapsto |\psi\left(\mathbf x\right)\rangle = \bigotimes_{j=1}^n \left( \cos\left(\frac{\pi}{2}x_j\right)|0\rangle - i\sin\left(\frac{\pi}{2}x_j\right)|1\rangle  \right),
\end{align}
with $n=k^2$.
}

\subsubsection{Threshold Encoding}
Another method of encoding classical data in quanvolutional filters is by \emph{threshold encoding}. It consists in first performing a binarization of the image, and then encoding pixels with value $0$ with the identity gate $I$, and pixels with value $1$ with the $X$ gate.
As in the previous case, the pixel $p_i$ is encoded in the $i$-th qubit.
This encoding process inevitably results in significant information loss due to the image binarization, and can only be applied to datasets that are resilient to this procedure.

It is important to note that the threshold encoding is equivalent to the rotational encoding after performing input binarization, as $R_X(\pi)$ is equivalent to the $X$ gate up to a global phase of $-i$.

\subsubsection{Higher Order Encoding}
The rotational encoding can be {enhanced} with additional entangling gates, to obtain the so-called \textit{higher order encoding}  \cite{HigherOrderEnc, Quanv_radiology}. In this encoding, after the rotational gates, there are a set of $R_{zz}(x_ix_j)$ applied to the $i$-th and $j$-th qubits. This encoding is more expressive, but requires additional $k^2\left(k^2-1\right)/2$ gates and a larger circuit depth.

\subsubsection{Other Notable Encodings}
Other encodings, which are usually not employed in quanvolutional approaches, aim to reduce the number of qubits needed to encode an image.
For examples Flexible Representation of Quantum Images (FRQI) \cite{FRQI} can encode an image of size $k \times k$ with $2\ log_2(k)+1$ qubits, as long as $k$ is a power of two. However, this method requires $k^4$ gates. Novel Enhanced Quantum Image Representation (NEQR) \cite{quanv_better_enc} is an improvement on the FRQI encoding that stores input data using the basis states instead of the amplitudes.

\subsection{Processing} \label{sec:henderson_processing}

Following the encoding phase, the processing section of the circuit usually consists of a randomly generated sequence of parametric and non-parametric gates.
In this section we focus on the original procedure described in \cite{Henderson2019QuanvolutionalNN}, as it is commonly used in the literature when implementing non-learnable \qnicks and it is the method used in this work to create the processing circuits.

The parametric circuit is constructed from a set of single-qubit gates, and a set of two-qubit gates.

The single-qubit gates are generated as follows: a maximum of $2k^2$ gates drawn from the set $[R_X(\theta), R_Y(\theta), R_Z(\theta), S, T, H]$. Each gate is applied to a random qubit, and $\theta$ is a random rotation parameter.

As regards the two-qubit gates, each pair of qubits $q_j, q_k$ has a fixed probability (usually $p=0.15$) of having a gate applied to them. The gate is randomly selected from the set $[CNot, Swap, SqrtSwap]$. The qubits selection is inspired by random graph models \cite{RandomGraphs}, where each qubit is treated as a vertex, and a two-qubit gate between them is treated as an edge. 

Finally, all the generated gates (single- and two-qubit) are randomly shuffled, obtaining the order of the gates for the final circuit $\mathcal U$.

\subsection{Decoding}

Finally, each quantum state must be translated into a classical, scalar value, in order to construct a new input for the following layer of the model. 
To achieve this, it is first necessary to measure each quantum state. Subsequently, the distribution obtained from the measurement can be converted into a real number. This step can be performed in several ways, e.g. the number of qubits in the $|1\rangle$ state in the most measured state can be counted \cite{Henderson2019QuanvolutionalNN}.
A different approach is the one used by \cite{Quanv_radiology}, where the authors obtain the expectation value for each observable in the circuit. Therefore, the output of the circuit is a vector instead of a single number, i.e. an output channel is generated for each qubit in the circuit.

\subsection{Expressibility of a Quantum Circuit}

Expressibility is one of the most significant descriptors of a parametric quantum circuit. It can be defined as the ability of the circuit to uniformly cover the Hilbert space of the underlying quantum system (i.e., the circuit's ability to explore the Bloch sphere in the case of a single qubit). Moreover, researchers have shown a strong positive correlation between the expressibility of a quantum circuit used in a variational quantum classifier and its performance \cite{exprCorrelation}.

The expressibility index has first been proposed in \cite{expressibility}. It is calculated by comparing the fidelities distribution of states obtained by a circuit $\mathcal U$ to the fidelities distribution of random states of the system, which corresponds to the Haar random states ensemble.

In order to compute the expressibility, the authors first approximate the former distribution by randomly sampling two sets of parameters $\mathbf\theta_1$, $\mathbf\theta_2\in\Theta$ of the parametric quantum circuit $\mathcal U$. They then compute the fidelity $|\langle \mathcal{U}(\mathbf\theta_1)|\mathcal{U}(\mathbf\theta_2)\rangle|^2$ between the states obtained with the corresponding sets of parameters. 

Subsequently, they compute the Kullback-Leibner divergence between the distribution $\Tilde{P}_\mathcal{U}(F;\Theta)$ and the distribution of random states $P_{\text{Haar}}(F)$, which is known to be equal to $(N-1)(1-F)^{N-2}$, where $N$ is the dimension of the quantum system \cite{AverageFidelity}, obtaining \begin{equation}
    Expr\left(\mathcal{U}(\mathbf{\theta})_{\mathbf{\theta\in\Theta}}\right) = D_{\text{KL}}\left(\Tilde{P}_{\mathcal{U}}(F;\Theta) \middle\|P_{\text{Haar}}(F)\right).
    \label{eq:expr}
\end{equation}

The formula to compute $D_{\text{KL}}$ of two continuous random variables $P$ and $Q$ is
\begin{equation}
    D_{\text{KL}}(P\|Q)=\int_{-\infty}^\infty p(x)\log\left(\frac{p(x)}{q(x)}\right)dx,
\end{equation}

where $p$ and $q$ denote the probability densities of $P$ and $Q$.

If the value of $D_\text{KL}$ (and therefore Expr) is close to zero, then the two distributions are similar (i.e. in our case, the parametric quantum circuit $\mathcal{U}$ is very expressive). 

It is important to note that in general expressibility is computed for variational circuits, while the circuits considered in this work have randomly generated or feature-dependent parameters.
In this context, it represents the ability of a circuit to extract diverse features from the input data.

\section{Proposed Method}
\subsection{Data Quantization}
\label{sec:quantization}

To enhance computational efficiency during the preprocessing stage, two techniques are employed: first the input data is quantized, and then quantum circuit outputs for each unique patch are memoized. Previous works utilized a binary image quantization and a look-up table (memoization) to expedite dataset processing. In general, binary quantization significantly reduces data fidelity, with the potential for complete loss of information (see Fig. \ref{fig:quantized_images}). Therefore, a higher number of quantization levels is employed to preserve of as much information as possible while maintaining computational practicality.
The proposed quantization approach extends binary quantization by introducing a quantization level, denoted as $N$.

The formula used to quantize a pixel is
\begin{equation}
    q(x) = \frac{\lfloor x\cdot N\rfloor }{N-1}
    \label{eq:quantization}
\end{equation}
where $x\in[0,1)$ is the original pixel intensity.
In other terms, when a image is quantized to $N$ levels, we first extract from the interval $[0,1]$ the $N$ points $\{0, \frac{1}{N-1}, \frac{2}{N-1}, \dots, 1\}$. Then, for each image $I$, each pixel value is mapped to the closest point in the set, obtaining a quantized image $q(I)$.

The memoization technique is implemented by constructing a look-up table that links each quantized input patch to the output computed by the circuit. The table is filled-in dynamically while preprocessing the dataset. This process ensures that unnecessary computations for absent patches are avoided, as only the patches actually encountered in the dataset are processed.

\Figure[htbp][width=0.99\linewidth]{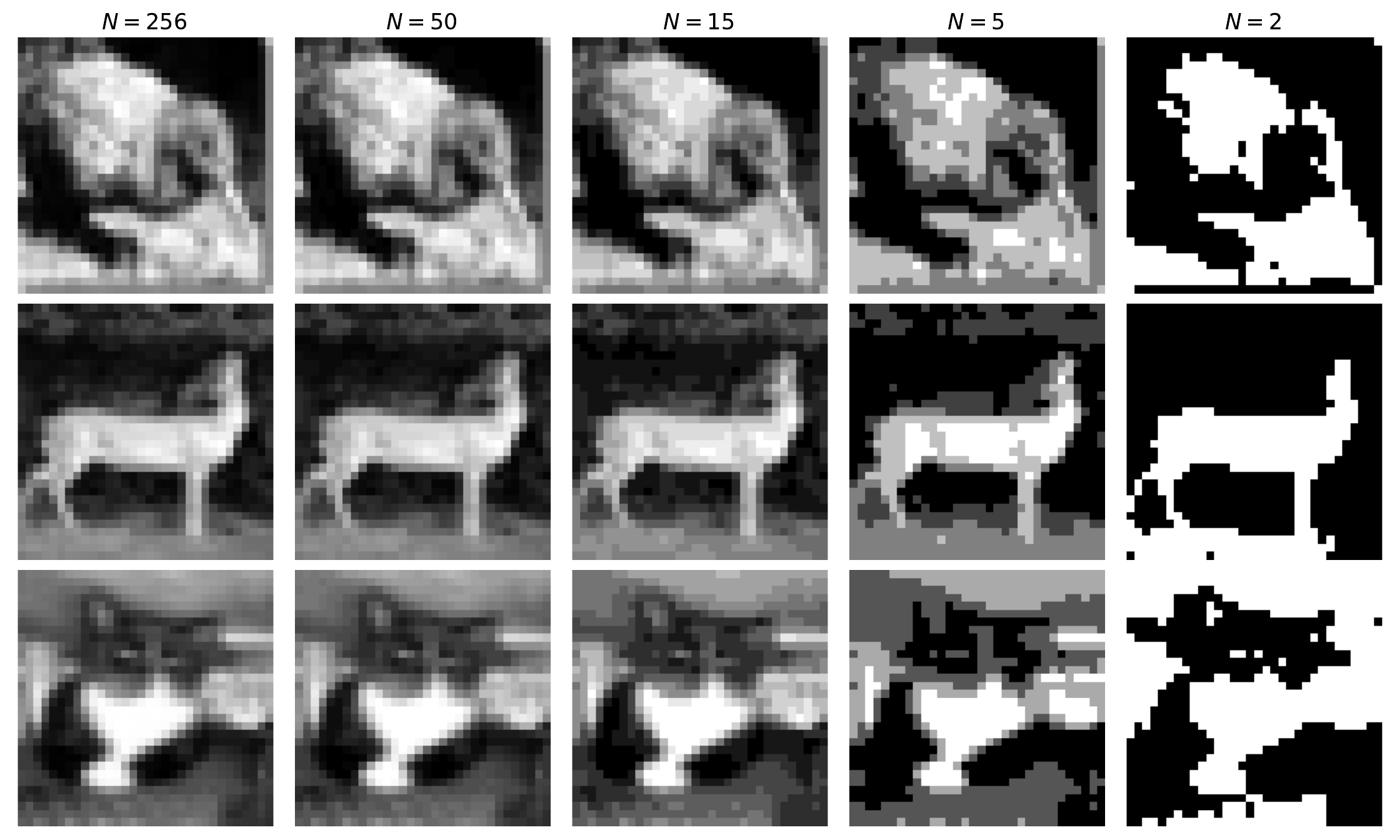}
{Visualization of quantization for different values of $N$, on three images from the CIFAR10 dataset (belonging to the classes bird, horse, and cat). Quantization is applied after converting to grey scale.\label{fig:quantized_images}}

This methodology allows to balance the trade-off between detail retention (preserving information) and computational load (efficient processing). A higher $N$ value offers finer details at the expense of expanding the memoization table and needing more quantum circuit executions. Conversely, a lower $N$ value simplifies the process and reduces computational demands, but this is achieved at the cost of losing input information. Fig. \ref{fig:quantized_images} displays some examples of the effect of quantization for different values of $N$.


To evaluate the information loss caused by using different quantization levels, we can calculate the mean squared error (MSE) between the original image and its quantized version. An upper limit can be estimated by considering that \eqref{eq:quantization} maps each value $x\in[0,1]$ to a point $q(x)$ such that $|x-q(x)|\leq \frac{1}{2\left(N-1\right)}$, therefore the error is 
\begin{equation}
\text{MSE}\left(I ; q(I)\right) \leq \frac{1}{4\left(N-1\right)^2}.
    \label{eq:max_mse}
\end{equation}

Nevertheless, the actual loss incurred during the quantization process ultimately depends on the specific dataset.


Without memoization, preprocessing a dataset would require a number of quantum circuit executions equal to the product of the number of input patches and the number of filters. Even for relatively modest datasets, this approach is impractical on current quantum hardware.
Previous methods relied on binarization (i.e. quantization for $N=2$) of the dataset to drastically reduce the amount of patches to compute (for example, the upper limit on the number of patches of size $3\times 3$ is $2^9=512$). However, larger values of $N$ might still provide a significant reduction in the number of unique patches, depending on the dataset. 

Since both MSE and the number of unique patches for a quantization level are computationally easy to obtain, these value can be used as indices to balance the information loss with the required processing time, without defaulting to an aggressive data binarization.

\subsection{Integrated Encoding}
\label{ref:integrated}
The proposed method  is based on the use of classical data as a rotation angle for multi-qubit entangling gates, with the objective of directly encoding data into the processing section of the circuit. The rationale behind this approach is that it removes the dependency between the size of the kernel and the nunber of qubits required, thus enabling their selection independently.
This approach is similar to some applications of quantum kernels \cite{kernel_methods}, where classical data is directly encoded in a quantum state without strict constraints on the number of qubits.

The parametric gates used in our model are obtained from the set of Pauli gates $P=\{\sigma_I, \sigma_X,\sigma_Y,\sigma_Z\} = \{I, X, Y, Z\}$. Given two elements $\sigma_1,\sigma_2\in P$, we can construct the parametric gate $G(\theta)$ as $\exp(-i \theta \sigma_1\otimes\sigma_2)$, where $\exp$ is the matrix exponential and $\otimes$ is the tensor product. As an example,  consider $\sigma_1=\sigma_2=X$. Then $\exp(-i\theta X\otimes X)=$ $$\begin{bmatrix}
\cos(\theta) & 0 & 0 & -i\sin(\theta) \\
0 & \cos(\theta) & -i\sin(\theta) & 0 \\
0 & -i\sin(\theta) & \cos(\theta) & 0 \\
-i\sin(\theta) & 0 & 0 & \cos(\theta)
\end{bmatrix}.
$$

Consider a filter of size $k\times k$, and a quantum circuit with $n$ qubits. A gate $G$ in the quantum circuit is constructed as follows:  a random feature $x_i \in \{x_1, x_2,\dots, x_{k^2}\}$ (corresponding to the intensity of a pixel in the patch $k\times k$) is selected, along with two gates $\sigma_1, \sigma_2 \in P$, two distinct qubits $q_j, q_k \in \{q_0,q_1,\dots, q_{n-1}\}$, and a mapping function $\alpha:[0,1]\to[0,2\pi]$. Then the gate $G$ is defined as
\begin{equation}
    G(x_i) = \exp\big(-i \alpha(x_i)\cdot\sigma_1\otimes\sigma_2\big)\big[q_j,q_k\big],
\end{equation}   
where the notation $[q_j,q_k]$ indicates that the gate is applied on qubits $q_j$ and $q_k$. The pair of qubits is randomly selected among all possible pairs. However, when working on a real quantum device, it is possible to base the selection on the connectivity of the device.
Finally, the mapping function $\alpha(\cdot)$ can be selected to be any mapping from the input space (i.e., pixel intensities) to rotation angles. In the simplest case, it can be chosen as $\alpha(x)=x\pi$. Nevertheless, it is possible to select alternative functions, such as random or learnable linear transformations, or standard activation functions such as the sigmoid function. 

The complete circuit can be written as a concatenation of $L$ gates $\prod_{l=1}^L G_l(x_{j_l})$.

To prevent information loss, we force the circuit generation to use each feature $x_i$ in the patch at least once.
The number of gates $L$ can be selected according to hardware limitations and desired properties, as long as $L\geq k^2$ to encode all the features. However, we expect the optimal number of gates $L$ to scale polynomially in both the number of features and the number of qubits $n$.

\subsubsection*{Comparison with Rotational Encoding}

One of the limitations of rotational encoding is its lack of flexibility.
In particular, a filter of size $k\times k$ requires exactly $k^2$ qubits,
and an average number of basic gates equal to $k^2+pk^4$,
where the former term depends on the single-qubit gates,
while the latter depends on the two-qubits connections between the $k^2$ qubits (see Section \ref{sec:henderson_processing}).
This results in circuits that may be excessively complex (in both depth or number of qubits)
when the size of the filter $k$ is high, or insufficiently expressive when $k$ is low. 

In contrast, our proposed model allows for significant flexibility in both the number of qubits and the number of operations $L$.

\begin{figure}[!hbtp]
\centering

\includegraphics[width=.9\linewidth, trim={0 3cm 27cm 0cm}, clip]{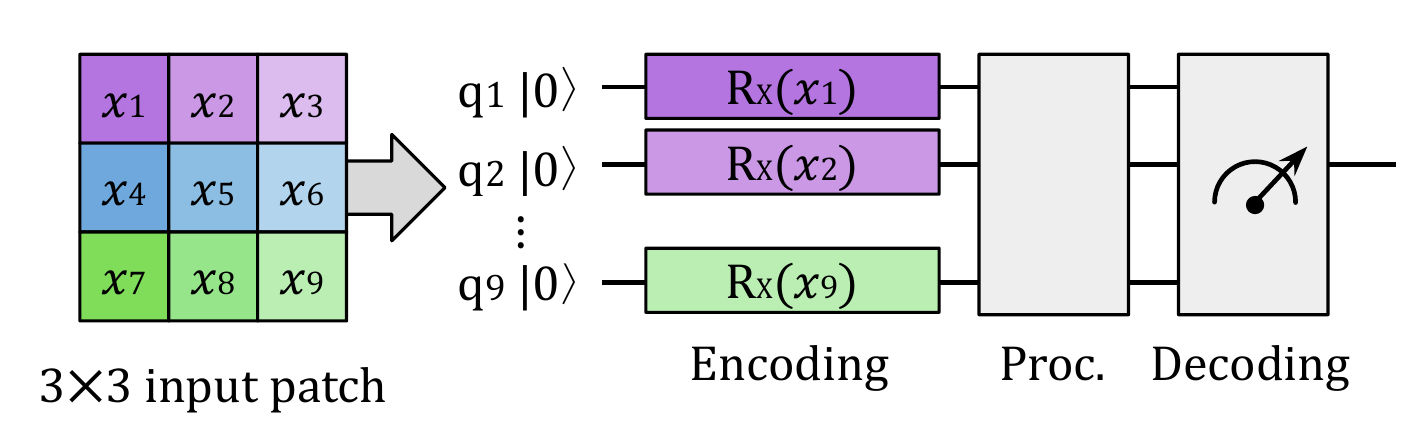}\\
(a)\\
\medskip
\includegraphics[width=.9\linewidth, trim={0 3cm 27cm 0cm}, clip]{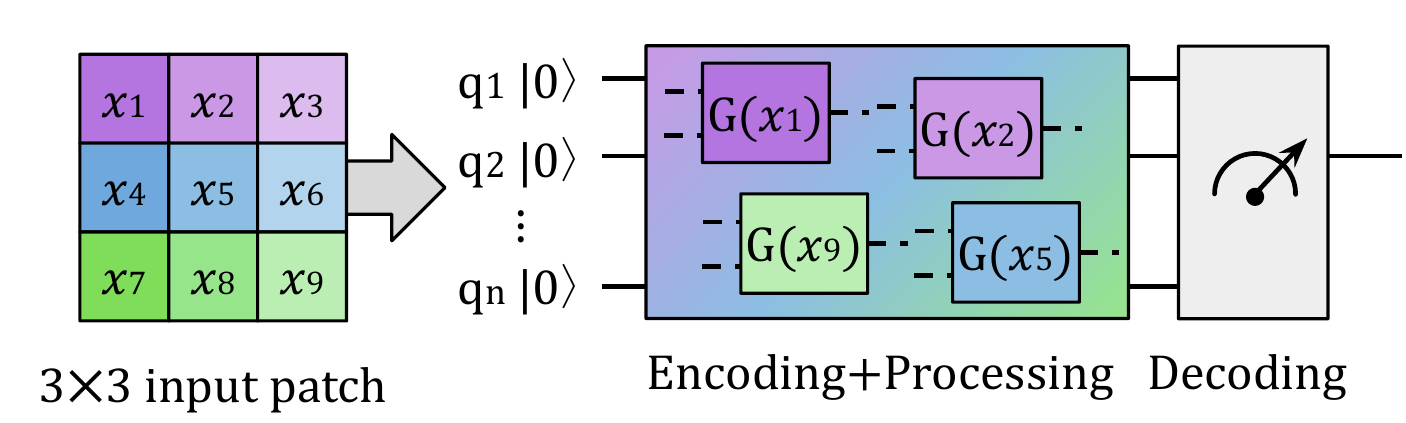}\\
(b)\\
\medskip

\caption{Graphical representation of the encoding methods compared in this work: (a) \encrotational, (b) integrated encoding.}
\label{fig:quanv_architectures}
\end{figure}

Fig. \ref{fig:quanv_architectures} shows a simplified comparison of the structure of a processing circuit using rotational encoding (a) and the proposed integrated encoding (b).

\section{Experimental Design}

The experiments are conducted on binary and multiclass classification problems across two different datasets. The aim of the experiments is to understand the performance of the proposed integrated quanvolutional model for different mapping functions $\alpha(\cdot)$, compared to the standard rotational encoding approach. Each setting is tested for different kernel sizes, and is finally compared to a classical CNN.  

\subsection{Datasets}

To test our proposed model, we selected two image classification datasets from different fields:

\begin{itemize}
    \item \textbf{\mirabest} \cite{mirabest_article} comprises images of galaxies, classified according to the Fanaroff–Riley morphology into three macro-classes: \textit{FR-I}, \textit{FR-II}, and \textit{Hybrid}. We use Version 1\footnote{Available at \underline{https://zenodo.org/records/4288837}} of the dataset, which consists of only the samples labeled as \textit{Confident} (as opposed to \textit{Uncertain}) and discards the Hybrid class, which contains 19 Confident samples only. The resulting dataset comprises $770$ samples, of which $339$ belong to the FR-I class and $431$ to the FR-II class. Each sample is normalized and down-scaled to a size of $30\times 30$. 

    \item \textbf{\lartpc} (Liquid Argon Time Projection Chamber) \cite{ChenHEPQCN} is a dataset consisting of realistic simulations of particle activities. Each particle belongs to a class in the following categories: $e^-, \mu^+, p, \gamma, \pi_0, \pi^+$ and $K^+$. Each class contains $100$ samples. Each sample is represented as a two-dimensional matrix, where one axis is the position in the sensing wire, the other axis is the time sampling tick, and the \quot{pixel intensity} is the energy loss at the corresponding position and time. The size of each matrix is $480 \times 600$. Samples are first normalized using a MinMax Scaler to ensure that each pixel is in $[0, 1]$, and then down-scaled to a size of $30\times 30$, following the original paper.

\end{itemize}

The main reasons for selecting these datasets are the following: first, their relatively small size ensures quicker processing times; second, the low amount of training samples allows us to test the capabilities of quantum models in low-resource settings. As shown in \cite{Caro2022}, quantum models can reach better generalization compared to classical models when few training data is available, making these datasets interesting comparison benchmarks.

As regards the \mirabest dataset, we used the fixed train-test split provided by the original authors ($90\%-10\%$), which results in $693$ train images and $77$ test images. For the \lartpc dataset, the data was randomly divided into a training set and a test set ($85\%$-$15\%$) at each experimental run.

\medskip

In addition, we employ the following datasets to test the effects of data quantization and analyze the trade-off between information loss and reduction in circuit executions:

\begin{itemize}
\item \textbf{MNIST} \cite{lecun2010mnist} comprises images of handwritten digits for 10-class digit classification. The dataset contains 70,000 images of size $28\times 28$.
\item \textbf{CIFAR10} \cite{Krizhevsky09learningmultiple} is a dataset for general image classification, containing 60,000 images of size $32\times 32$ belonging to 10 classes of vehicles and animals.

\end{itemize}

Table \ref{table:dataset_size} summarizes the basic information about the datasets, such as the number of samples, image size, and the total number of unique $3 \times 3$ patches.

\begin{table}[t!]
\caption{Number of samples and sample size for four image classification datasets:  MNIST, CIFAR10, \mirabest, and \lartpc. Patches of size $3\times 3$ are used for illustrative purposes.}
\centering
\begin{tabular}{rrrr}
\hline
\shortstack{Dataset\vspace{.2em}\\ \  } & \shortstack{Samples\\ \ } & \shortstack{Image Size\\ \ } & \shortstack{\\Number of\\ $3\times 3$ patches} \\
\hline
MNIST     & 70,000 & {$28\times 28$} & 47,320,000 \\
CIFAR10   & 60,000 & {$32\times 32$} & 54,000,000 \\
\mirabest &    770 & {$30\times 30$} &    603,680 \\
\lartpc   &    700 & {$30\times 30$} &    548,800 \\
\hline
\end{tabular}
\label{table:dataset_size}
\end{table}


\subsection{Tested Models}

We compare our proposed quanvolutional model with integrated encoding (\qnnintegrated) with a classical CNN model and a quanvolutional network using rotational encoding (\qnnrotational). Additionally, we test three different mapping functions $\alpha(\cdot)$ to map pixel intesities to rotations angles in \qnnintegrated. Finally, for both quanvolutional models, multiple kernel sizes are tested to determine their effect on the model's performance. Following are the implementation details of the models. Fig. \ref{fig:architecture} shows a schema of the architecture.

\Figure[htbp](topskip=0pt, botskip=0pt, midskip=0pt)[width=\linewidth]{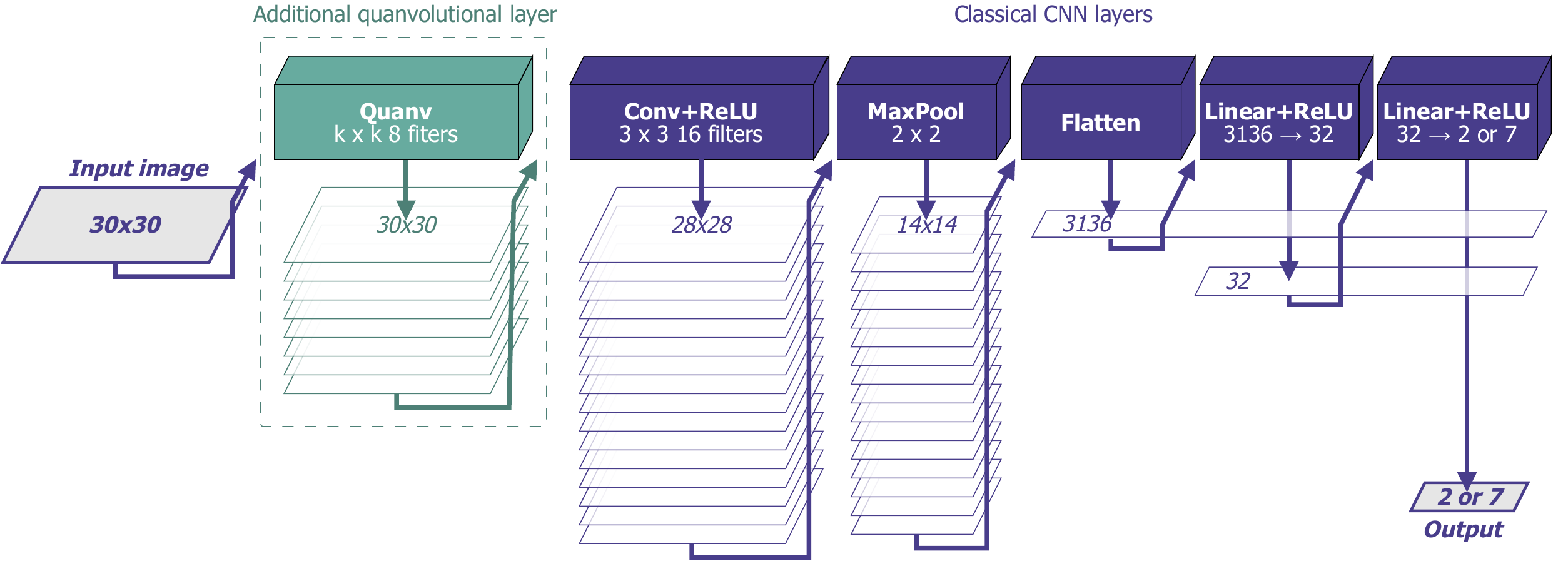}
{Schema of the CNN and QNN-based architectures. The QNN-based architectures are obtained by adding a quanvolutional layer (in green) before the classical CNN layers (in blue).\label{fig:architecture}}

\begin{itemize}
\item {CNN}: a classical convolutional neural network. The structure of the model contains one convolutional layer and two fully connected layers. The convolutional layer consists of $16$ output channels, has a filter size of $3\times 3$ with no padding, and is followed by a ReLU, and a Max Pooling layer of size $2\times 2$. The first fully connected layer has $32$ output features, while the second one, which is also the output layer, has a number of outputs equal to $7$ or $2$, depending on the task. Each fully connected layer is followed by a ReLU activation function. The convolutional layer and the first fully connected layer are followed by a Droupout layer, with probability of $0.2$, to prevent overfitting.

\item {\qnnrotational}: quanvolutional model with rotational encoding. The model consists in the same architecture described above, with a quanvolutional layer stacked on top. The number of output channels on the quanvolutional layer is $8$, and the first Conv layer is modified is take as input $8$ channels instead of one. Each filter is padded to have the same input and output dimensions.

The tested filter sizes are: $2\times 2$, $3\times 3$, and $4\times 4$, with a qubit requirement $n$ of $4, 9$, and $16$. Larger filter sizes were not tested due to the high resource demands associated with their simulation, which was unfeasible on our device. The processing circuit follows the standard implementation describe in Section \ref{sec:henderson_processing}. The connection probability used for the circuit generation is set to $0.15$.

\item {\qnnintegrated}: quanvolutional model using the proposed integrated circuit. The structure of the model is the same as the one described for the QNN with the rotational encoding, with the quanvolutional layer on top of the CNN. The layer comprises $8$ channels, and each filter is padded so that the output of the layer has the same dimensions of the input.

The number of qubits is set to $n=4$ to balance the expressivity of the model with the resources required for its implementation. The tested filter sizes are: $2\times 2, 3\times 3, 4\times 4$ and $5\times 5$.

Regarding the mapping function $\alpha(\cdot)$, three options were considered:
\begin{itemize}
    \item \SIMPLE:  $x\mapsto x\pi$ 
    \item \RNDMUL: $x\mapsto 2\beta x \pi$ 
    \item \RNDLIN: $x\mapsto (\beta x + \sigma)\pi$  
\end{itemize}
where $x$ is the pixel value normalized in $[0, 1]$, and $\beta, \sigma$ are random parameters selected uniformly in $[0, 1]$, different for each gate. 

As discussed in Section \ref{ref:integrated}, the number number of gates $L$ should be at least equal to $k^2$ to encode all input features. We set $L=2k^2$, and force each feature to be encoded at least once in the final circuit.

\end{itemize}

\subsection{Decoding}

To obtain the output of the quanvolutional filter, we need a way to decode the quantum state through a measurement. In this work, we choose to measure the output state $\mathcal{U}(\mathbf{x})|\psi_0\rangle$ with the projector $\mathcal M = Z^{\otimes n}$. The expectation value is obtained as \begin{equation}
    p = \langle\psi_0|\mathcal{U}(\mathbf{x})^\dagger\mathcal{M}\mathcal{U}(\mathbf{x})|\psi_0\rangle.
\end{equation}

To translate the output measurement to a scalar value, we compute the average number of qubits that were measured in the $|1\rangle$ state after repeating the measurement for a fixed number of times.
This method was chosen as it is more resilient to the stochasticity of the measurements than the one based only on the most common measured output \cite{Henderson2019QuanvolutionalNN}.
As an example, consider a circuit that for a given input patch $\mathbf{x}$ returns an equal superposition of the states $|0\rangle^{\otimes n}$ and  $|1\rangle^{\otimes n}$. If the filter output depends only on the most common state measured, then the output has a $50\%$ chance of being $0$ or $1$, while by taking into account all the measured states we have an output of $0.5$.

\subsection{Implementation Details}

All models are trained using the ADAM optimizer with a learning rate of 0.0003, with a batch size of $16$. As loss, we use log softmax.
The training process employs early stopping with a patience of $10$, i.e. the training halts if the training loss fails to improve over $10$ consecutive epochs. For the \qnnintegratedRNDLIN model the patience is increased to $100$, as the training loss took significantly more epochs to decrease.
Each training is repeated $10$ times with different random seeds.

The model architectures, training procedures and testing methodologies are implemented in PyTorch.

Quantum operations are performed with noiseless simulation with Qiskit 1.1. The decoding of the circuits is performed by a standard measurement in the computational basis for each qubit, and by computing the average (i.e. the final output is the average number of qubits in the state $|1\rangle$) for a fixed number of samples equal to $1000$. In general, by increasing the number of qubits in the system, it is expected that (exponentially) more measurements are required to obtain a good estimation of the quantum state. However, since the number of qubits used in the experiments is small, we did not observe any significant difference when increasing the number of shots.  

\subsection{Expressibility Measurement}

To compute expressibility of a quanvolutional circuit, we compute $2^{10}$ fidelities of the circuit by creating pairs of random input vectors. We then calculate the discretized version of Expr as follows:

\begin{equation}
    \sum_{i\in\text{bins}}P(i)\log \left(\frac{P(i)}{Q(i)+\varepsilon}\right),
\end{equation} 

where bins are obtained by dividing the interval $[0,1]$ in $50$ equal-sized intervals, $P(i)$ is the number of fidelities in the $i$-th bin, and $Q(i)$ is obtained from the distribution of $P_{\text{Haar}}$. The additive constant $\varepsilon=10^{-16}$ is used for numerical stability in the computation.

All the randomly initialized non-learnable parameters of the circuits (e.g. $\theta$ in the \qnnrotational, $\sigma,\beta$ in \qnnintegrated) are maintained fixed while measuring the $2^{10}$ fidelities.

The expressibility measurement described above is repeated 10 times for differently initialized circuits.

\section{Experimental Results}

In the following, we present an analysis of the effect of data quantization on the datasets. Then, we report the results of the classification experiments conducted on \mirabest and \lartpc. Finally, we perform an expressibility analysis for \qnnintegrated and \qnnrotational.

\subsection{Data Quantization}

We perform a preliminary analysis of the impact of data quantization, restricted to $3\times 3$ patches on the four datasets: MNIST, CIFAR10, \mirabest, and \lartpc. We focus on two metrics: information loss and reduction in number of circuit execution. The former is calculated as the MSE between the original and quantized images. The latter is calculated by comparing the total number of $3\times 3$ patches in the whole dataset and the number of unique $3\times 3$ patches obtained after quantization.


\medskip

Fig. \ref{fig:quantization_error_log} shows the trend of the information loss (MSE, y-axis) depending on the quantization level (x-axis) for all datasets.
The theoretical maximum MSE between the original and quantized images for each quantization level $N$, given by \eqref{eq:max_mse}, is shown as a dashed line.
The actual average MSE calculated on the datasets is shown as solid lines.
We can see that the actual MSE on the datasets can be significantly lower than the upperbound (e.g., for \lartpc and \mirabest). This largely depends on the variability of pixel intensities in the original dataset. For example, CIFAR10 contains real-life images of animals, vehicles, and objects (see Fig. \ref{fig:quantized_images}), resulting in a high variability and an MSE close to the upper bound. On the other hand, the images in \lartpc and \mirabest contain large black backgrounds with (relatively) small mostly-white objects, which lead to lower error rates during quantization.

The acceptability of information loss is contingent upon the characteristics of the dataset in question. For instance, $N=2$ is an acceptable quantization level for MNIST, whose images remain recognizable after the process, but not for CIFAR10. In general, we expect that levels of $N\ll 20$ may result in a loss of information that negates any potential quantum advantage that could be obtained through quanvolutional approaches.

\Figure[htbp](topskip=0pt, botskip=0pt, midskip=0pt)[width=.99\linewidth,]{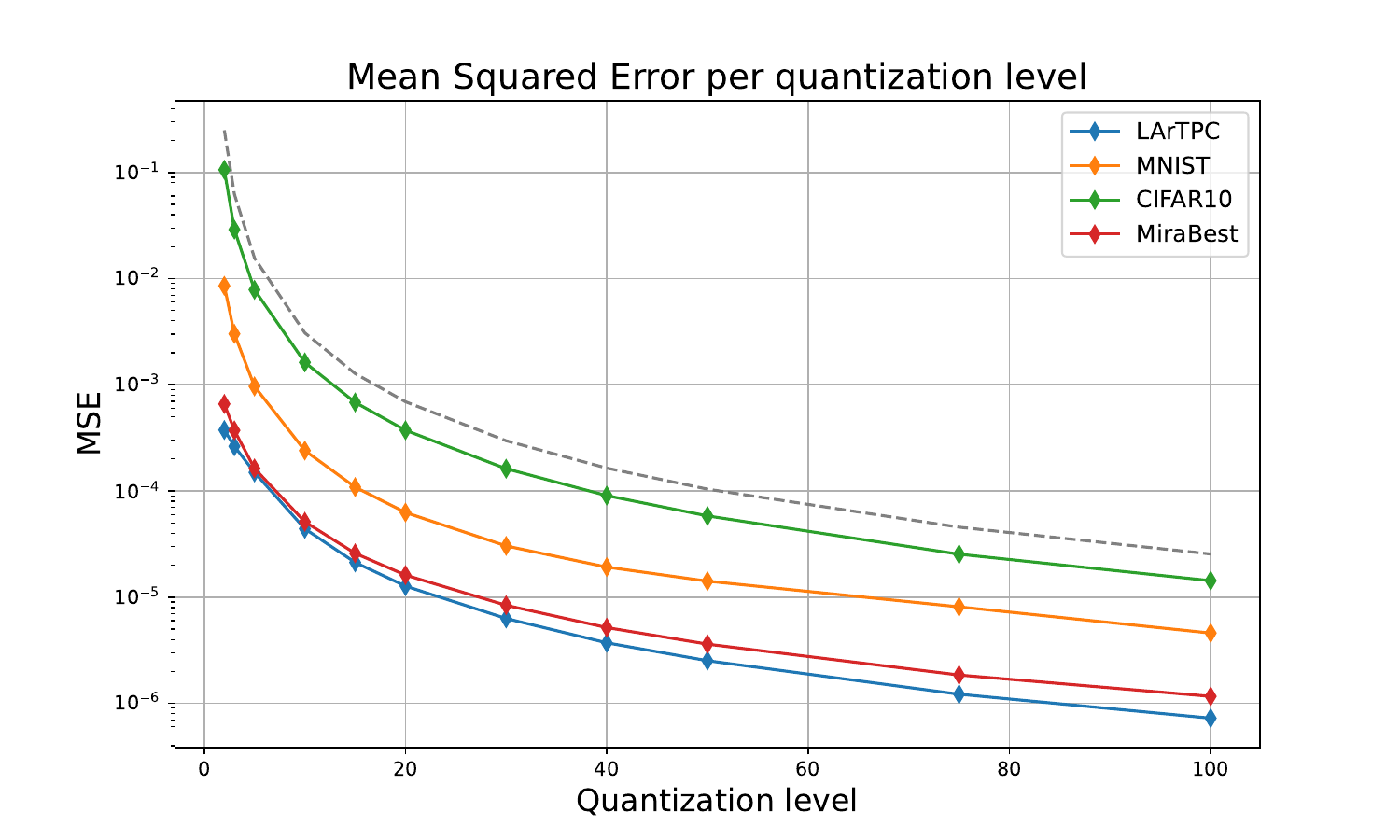}
{Average MSE on the considered datasets after applying quantization to $N$ levels, shown on a logarithmic scale. The dashed line is the maximal MSE given by \eqref{eq:max_mse}.\label{fig:quantization_error_log}}

\medskip

Fig. \ref{fig:reduction_circuit_exec} reports the percentage of reduction in circuit executions (y-axis) given the quantization level (x-axis) for all datasets.
The plot shows that quantizing to $N=10$ levels reduces the number circuit executions by 92\% for all datasets. The reduction reaches 99\% for \mirabest and \lartpc, potentially allowing significant savings in terms of resource utilization.
As the number of quantization levels increases, the effect of quantization remains significant for several datasets: $N=100$ leads to a 95\% reduction for \mirabest and \lartpc, and an 80\% reduction for MNIST.
On the other hand, we observe that for CIFAR10 the amount of reduction in circuit executions becomes minor for $N\gg20$, reaching 43\%, 24\%, and 10\% for $N=30, 50,$ and $100$ respectively.

This confirms the previous observations on the information loss, showing that the efficacy of this method ultimately depends on the characteristics of the dataset.

Given these results, the \mirabest and \lartpc datasets used in the following experiments are preprocessed with $N=50$ quantization levels. The value of $N$ is selected to obtain a high computational speed-up in the quanvolutional layer application (respectively 96\% and 97\% reduction in circuit executions, see Table \ref{tab:quantization_reduction}) with minimal information loss (MSE$<10^{-5}$, see Fig. \ref{fig:quantization_error_log}).

\begin{table*}[htbp]
\centering
\caption{Reduction in the number of circuit executions after applying quantization ($N$ levels) and memoization to the datasets \lartpc, \mirabest, and MNIST with patch size $3\times3$. In bold, the level selected for the experiments in this work.}
\begin{tabular}{l rrrrrrr}
\hline
Quantization Level & 5 & 10 & 15 & 20 & 30 & \textbf{50} & 100 \\
\hline
LArTPC & 99.91 & 99.51 & 99.07 & 98.68 & 98.04 &\textbf{ 97.07} & 95.79\\
MNIST & 99.64 & 97.29 & 94.51 & 92.10 & 88.57 & 84.66 & 80.99\\
CIFAR10 & 99.62 & 92.90 & 78.68 & 64.18 & 43.40 & 23.82 & 9.93\\
MiraBest & 99.79 & 99.08 & 98.48 & 98.03 & 97.39 & \textbf{96.61} & 95.56\\
\hline
\end{tabular}
\label{tab:quantization_reduction}
\end{table*}

\Figure[t!](topskip=0pt, botskip=0pt, midskip=0pt)[width=.99\linewidth,]{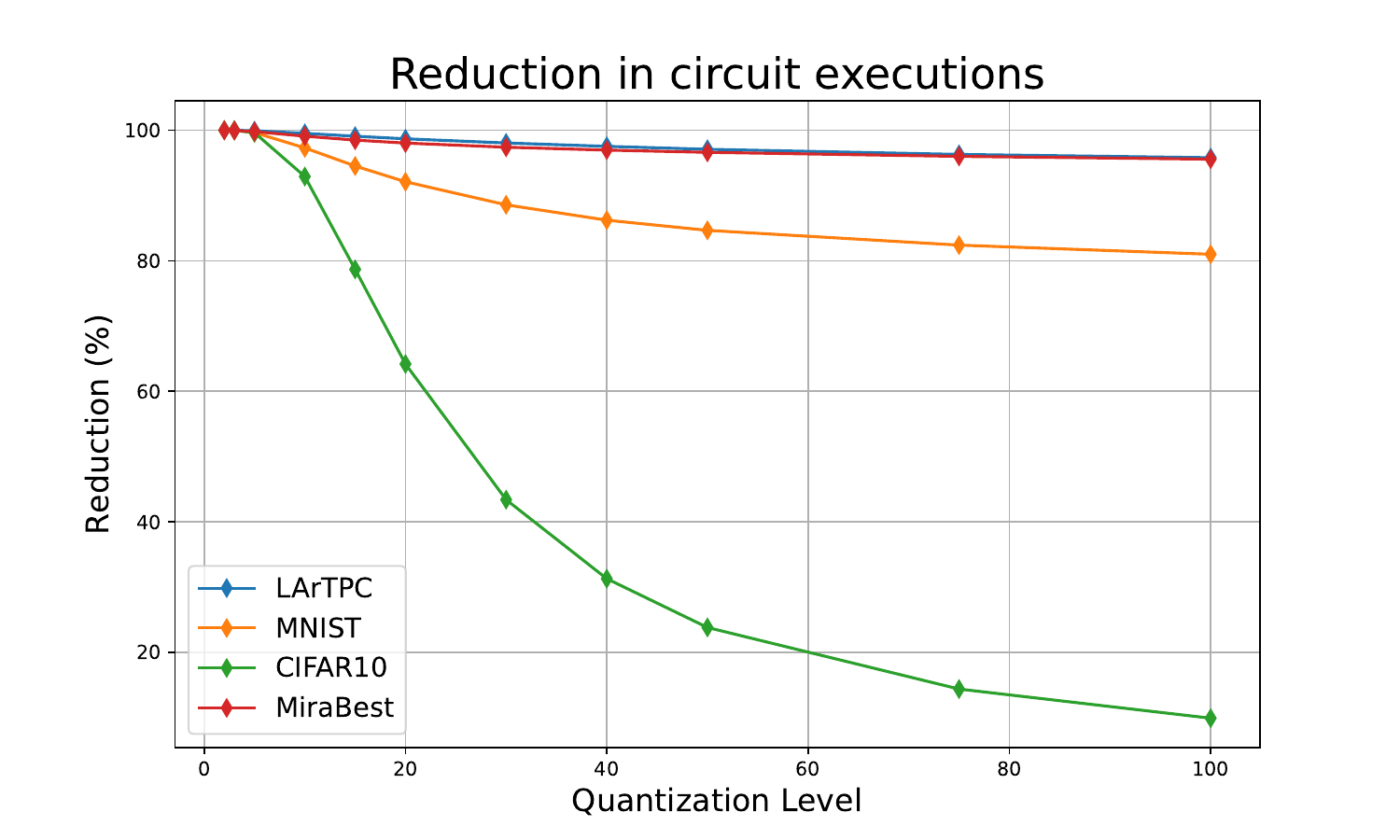}
{Reduction in the number of circuit executions after applying quantization ($N$ levels) and memoization to the considered datasets with patch size $3\times3$.\label{fig:reduction_circuit_exec}}

\subsection{Classification performance}

\subsubsection{Results on \mirabest}

Table \ref{tab:mirabest_data} reports the average classification accuracy of all models tested on the \mirabest dataset (two classes).

The average accuracy of the classical CNN model is 71.039\%.

\qnnrotational surpasses the performance of the CNN for $k=2,3$, reaching up to 77.273\% accuracy. Its performance degrades as $k$ increases, and becomes unstable with $k=4$, with 10 points of standard deviation.

\qnnintegratedSIMPLE and \RNDMUL significantly outperform both the baseline CNN and \qnnrotational for $k=2,3,4$, and \RNDMUL also matches the performance of the CNN with $k=5$. \qnnintegratedRNDMUL reaches the best overall performance with 80.779\% accuracy at $k=3$.

The \qnnintegratedRNDLIN model exhibits sub-optimal performance for all values of $k$, as its performance remains 10--20 points lower than the classical CNN.

In general, lower values of $k$ tend to perform better, while there is a decline in performance as $k$ increases to 4 and 5.

\begin{table*}[ht]
\caption{Classification accuracy for the \mirabest dataset. The reported results are the average and standard deviation over 10 runs. Numbers in bold correspond to values higher than the baseline CNN accuracy.}
\centering
\begin{tabular}{l rrrr}
\hline
Model & {$k=2$} & {$k=3$} & {$k=4$} & {$k=5$} \\
\hline
\qnnrotational & {\textbf{77.273} $\pm$ 2.740} & {\textbf{75.584} $\pm$ 2.309} & {67.922} $\pm$  9.993& - \\
\qnnintegratedSIMPLE & {\textbf{79.870} $\pm$  2.126} & {\textbf{79.870} $\pm$ 2.548} & \textbf{{74.156}} $\pm$  6.859& {69.351} $\pm$  6.118\\
\qnnintegratedRNDMUL & {\textbf{80.130}} $\pm$  2.098& {\textbf{80.779}} $\pm$ 2.158 & {\textbf{73.247}} $\pm$  7.731& 71.039 $\pm$  5.896\\
\qnnintegratedRNDLIN & {63.636} $\pm$  9.719& {61.492} $\pm$  8.771& {57.792} $\pm$  3.907& {58.052} $\pm$  6.623\\
\hline
\multicolumn{1}{c}{CNN} & \multicolumn{4}{c}{{71.039 $\pm$ 12.632}} \\
\hline
\end{tabular}
\label{tab:mirabest_data}
\end{table*}

\subsubsection{Results on \lartpc}

Table \ref{tab:LArTPC_data} reports the average classification accuracy of all models tested on the \lartpc dataset (seven classes).

The average accuracy of the classical CNN model is 56.789\%.

The performance of the \qnnrotational model is lower than the CNN for all values of $k$. Its performance increases with $k$, and it reaches a maximum of 52.929\% for $k=4$.

\qnnintegratedSIMPLE and \qnnintegratedRNDMUL have similar patterns in performance. For $k=2$, their accuracy is lower than the CNN, but still higher than the best performance of \qnnrotational. For $k=3,4,5$ both models surpass the performance of the CNN, reaching their best performance at $k=4$ (58.500 for \SIMPLE and 57.786 for \RNDMUL).

\qnnintegratedRNDLIN never surpasses the baseline CNN performance, reaching a maximum of 54.071\% accuracy with $k=2$, and showing a large standard deviation for $k=3,4,5$ (up to 18 points). This shows that \RNDLIN is highly unstable on this dataset.

In contrast to the \mirabest dataset, larger filters obtain better results, but the best performing models exhibit a smaller advantage over the classical approach. This might be caused the higher number of output classes in the dataset.

\begin{table*}[ht]
\caption{Classification accuracy for the \lartpc dataset. The reported results are the average and standard deviation over 10 runs. Numbers in bold correspond to values higher than the baseline CNN accuracy.}
\centering
\begin{tabular}{l rrrr}
\hline
Model & {$k=2$} & {$k=3$} & {$k=4$} & {$k=5$} \\
\hline
\qnnrotational & {51.571} $\pm$  3.270 & {52.357} $\pm$ 3.900 & {52.929} $\pm$  4.354& - \\
\qnnintegratedSIMPLE & {54.143} $\pm$ 3.044& {\textbf{56.857}} $\pm$  3.067& {\textbf{58.500}} $\pm$  2.246& {\textbf{58.429}} $\pm$  2.299\\
\qnnintegratedRNDMUL & {54.071} $\pm$ 2.434 & {\textbf{57.643}} $\pm$  3.147& {\textbf{57.786}} $\pm$ 2.106 & {\textbf{57.786}} $\pm$  2.565\\
\qnnintegratedRNDLIN & {54.071} $\pm$  3.572& {45.786} $\pm$ 16.368 & {49.571} $\pm$  18.237& {51.286} $\pm$  13.328\\
\hline
\multicolumn{1}{c}{CNN} & \multicolumn{4}{c}{{56.786} $\pm$  9.182}\\
\hline
\end{tabular}
\label{tab:LArTPC_data}
\end{table*}

\subsection{Expressibility Analysis}

\begin{figure*}[htbp]
    \centering
    
    \includegraphics[height=11em, trim={0cm .2cm 0cm 0cm}, clip]{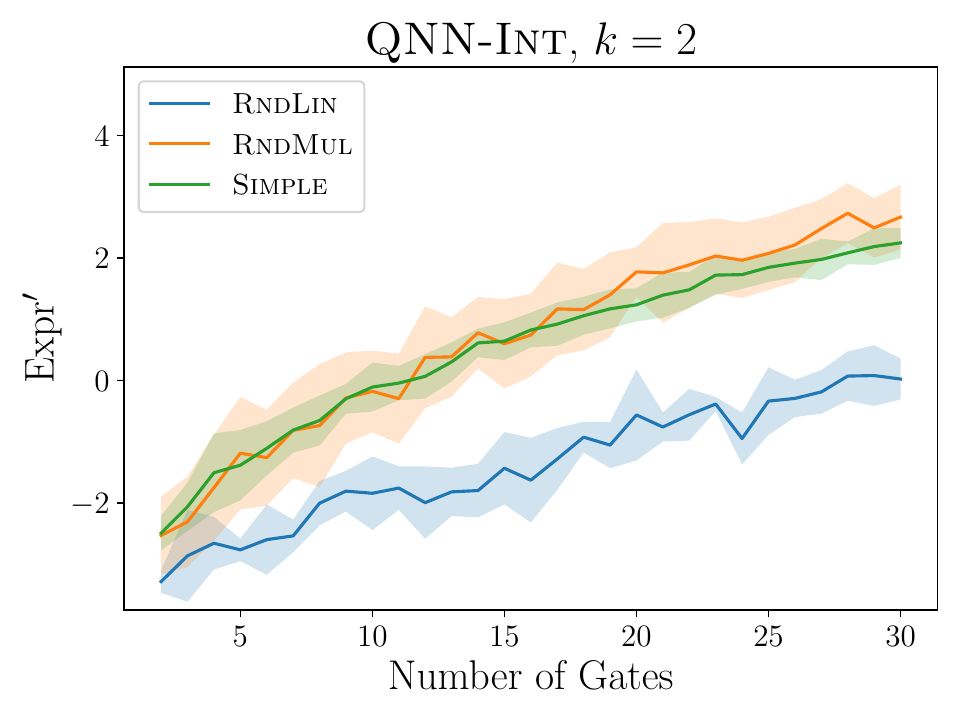}
    \includegraphics[height=11em, trim={1.5cm .2cm 0cm 0cm}, clip]{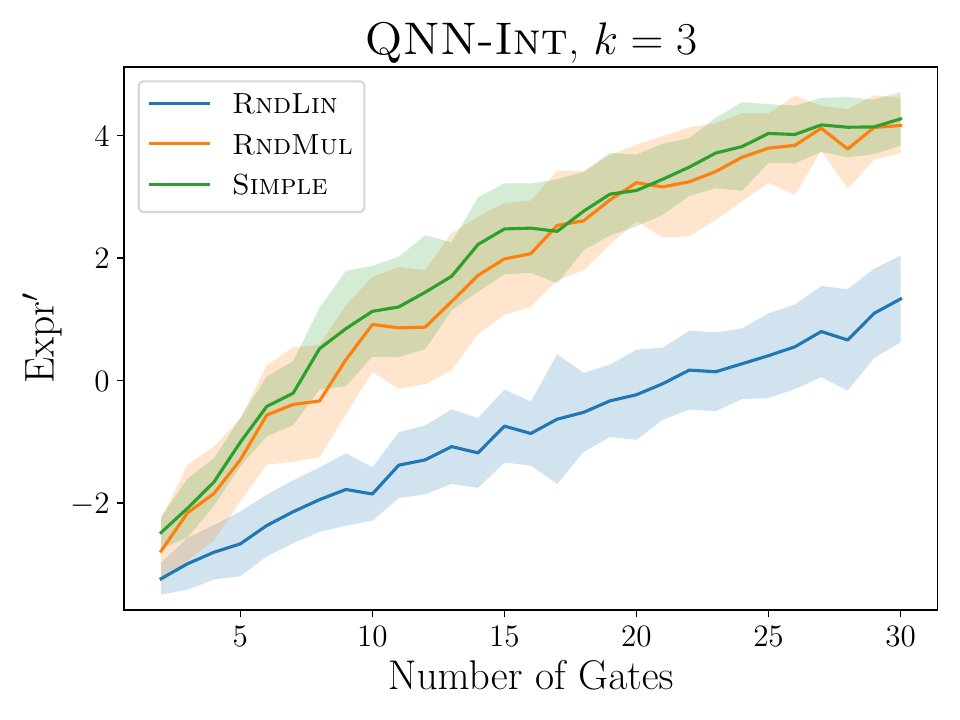}
    \includegraphics[height=11em, trim={1.5cm .2cm 0cm 0cm}, clip]{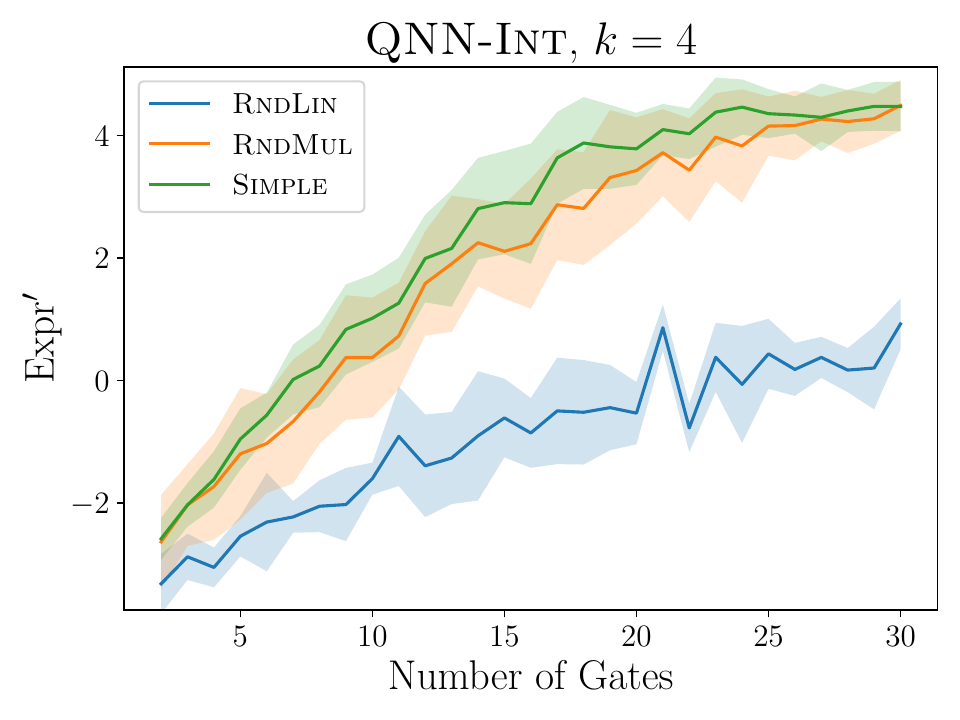}\\ \medskip
    \includegraphics[height=11em, trim={0cm .2cm 0cm 0cm}, clip]{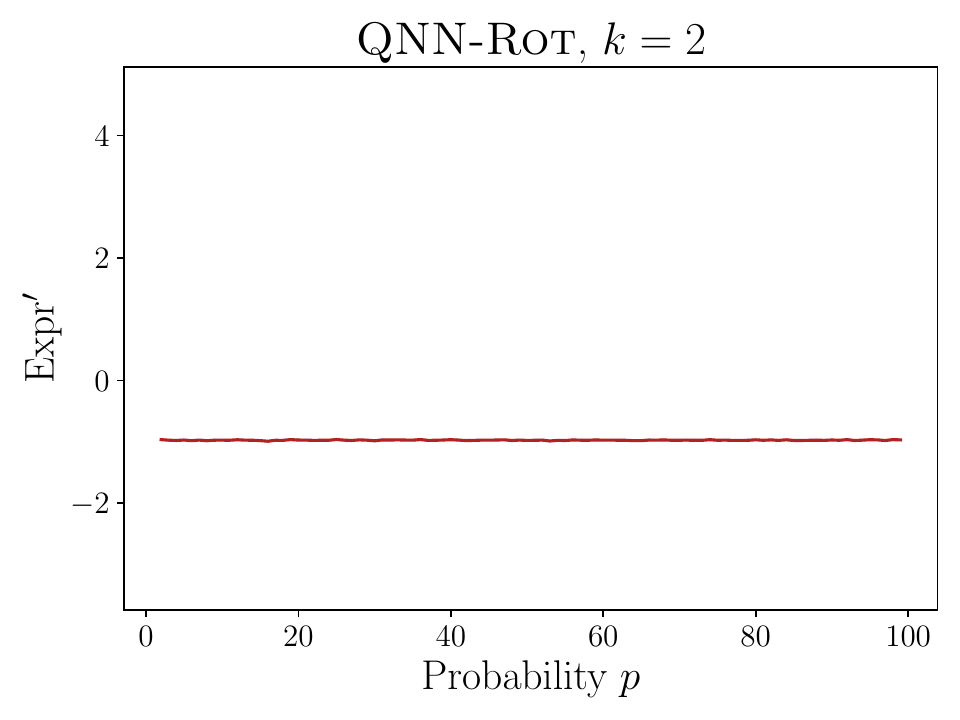}
    \includegraphics[height=11em, trim={1.5cm .2cm 0cm 0cm}, clip]{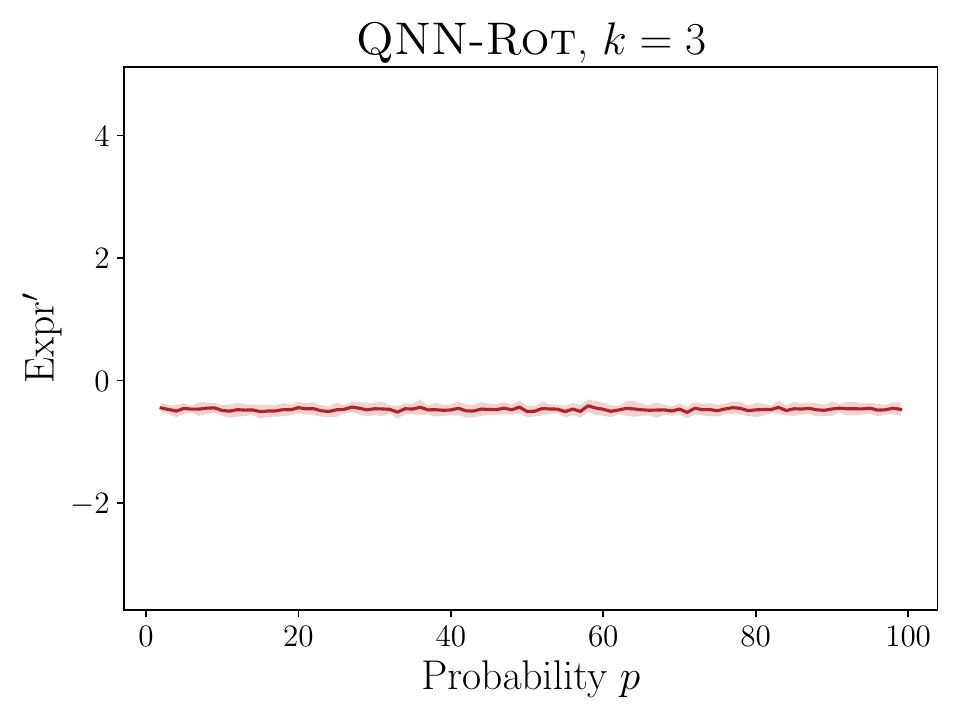}
    \includegraphics[height=11em, trim={1.5cm .2cm 0cm 0cm}, clip]{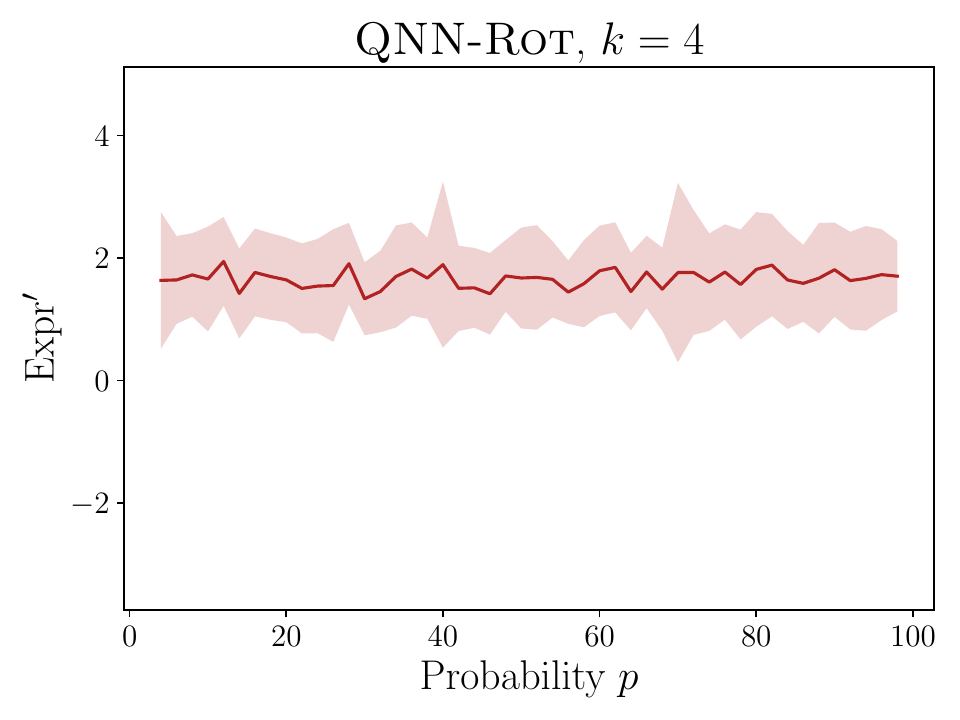}
        
    \caption{Expressibility of parametric quantum circuits, computed as described in Formula \ref{eq:expr} by uniformly sampling parameters in $[0,1]$. For readability purposes, we show the value Expr$'$, computed as $\text{Expr}'=-\ln(\text{Expr})$. For the model \qnnintegrated, the number of qubits is set to $4$. All plots report the average expressivity (solid line) and its standard deviation (shadowed area) over 10 random circuits.}
    \label{fig:expressibility}
\end{figure*}

Fig. \ref{fig:expressibility} shows a comparison of the expressibility of the proposed \qnnintegrated model with \qnnrotational using different kernel sizes.
To increase the readability of the plots, we show the value of Expr$'$, computed as $\text{Expr}'=-\ln(\text{Expr})$, so that higher values correspond to an increased expressivity.

When computing the expressibility of the proposed \qnnintegrated circuit with a fixed number of qubits and features (Figs. \ref{fig:expressibility}, upper row), we observe that increasing the number of gates $L$ (moving right on the x-axis) leads to increasingly more expressive circuits (increased Expr$'$). This implies that it is possible to select \textit{a priori} the value of $L$ in order to obtain a desired value of expressibility.

Increasing the kernel size $k$ leads to a higher increase in expressibility with a lower number of gates. For example, \qnnintegratedSIMPLE reaches Expr$'=2$ with 25 gates when $k=2$, 14 gates with $k=3$, and 12 gates with $k=4$.

When comparing the three different mapping functions $\alpha$, we observe that \SIMPLE and \RNDMUL display a similar trend, while \RNDLIN has significantly lower expressibility for all kernel sizes. Therefore, we expect \RNDLIN to reach a lower classification accuracy compared to the other two functions.

As regards \qnnrotational, the x-axis of the plots in the lower row of Fig. \ref{fig:expressibility} reports the probability $p$ instead of the number of gates. This is because the number of gates in a circuit with rotational encoding depends on $p$ and $k$ and it averages at $k^2+pk^2(k^2-1)$. One can select different values of the connection probability $p$ to obtain circuits with different length.
We observe that \qnnrotational circuits are generally less expressive than \qnnintegrated. Additionally, their expressibility does not depend on the number of operations involved. For every value of $p$ there is no statistically significant difference in the value of Expr$'$, which only increases with the kernel size $k$.

These observations are in line with the experimental results, which show that \qnnintegratedSIMPLE and \qnnintegratedRNDMUL tend to outperform \qnnrotational in classification tasks.

\subsection{Discussion}

Overall, the results of the classification experiments on both datasets show that the proposed \qnnintegrated model reliably surpasses the performance of the \qnnrotational model for all kernel sizes, and was always able to surpass the performance of a classical CNN.

In the \qnnintegrated model, the function chosen to map pixel intensities to rotation angles plays an important role and can lead to very different performances. Among the tested functions, \SIMPLE and \RNDMUL performed the best, while \RNDLIN often reached lower performances even compared to \qnnrotational. This behaviour could be linked to the lower expressibility of \RNDLIN.

Increasing the kernel size $k$ led to different results in the two datasets. On \mirabest, the performance of the models seems to decrease with $k$, while on \lartpc the best results are achieved for higher values of $k$. This phenomenon might be linked with the type of images in the dataset, the complexity of the task, or the number or output classes, and it deserves further exploration.

\section{Conclusions}

In this work, we presented a new quanvolutional model and preprocessing pipeline to make data quantization, encoding, and processing more efficient on NISQ devices.

The proposed flexible quantization approach enabled a significant reduction in the number of quantum circuit executions required to process the datasets considered in this work. In particular, we obtained a reduction of over $95\%$ circuit executions when using $3\times3$ kernels, while losing a negligible amount of information. This technique has the potential to be highly beneficial for quanvolutional approaches applied to tasks with similar properties.

Our experiments also show that the proposed \qnnintegrated model can match or surpass the performance of classical CNN models on different datasets and with different parameter configurations.
When compared with a standard quanvolutional model with rotational encoding (\qnnrotational), \qnnintegratedSIMPLE and \qnnintegratedRNDMUL surpassed its performance on all tested configurations. 

The proposed integrated encoding model features a large number of hyperparameters, including the number of qubits $n$, the filter size $k$, and the number of gates $L$, in addition to an extensive range of possible mapping functions $\alpha(\cdot)$. Each parameter can be selected independently based on the hardware constraints of current quantum devices. The choice of the mapping function seems to be crucial to ensure higher expressivity and better classification results. 

In the future, it would be interesting to test the proposed integrated encoding in a {learnable} fashion. In particular, mapping functions such as the ones used in \qnnintegratedRNDLIN and \qnnintegratedRNDMUL have randomly initialized parameters which could be optimized during training, as done in \cite{quanv_better_enc} for \qnnrotational.

\section*{Data Availability}

The code to generate and test quanvolutional models, with both rotational encoding and integrated model, can be obtained at \underline{https://github.com/Dan-LB/integrated\_}\break\underline{encoding\_for\_QuanvNN}. 
The \mirabest dataset can be downloaded at \cite{mirabest_dataset}. The \lartpc dataset can be obtained by requesting it to the authors of \cite{ChenHEPQCN}.

\bibliographystyle{IEEEtran}
\bibliography{bibliography}

\newpage

\begin{IEEEbiography}[{\includegraphics[width=1in,height=1.25in,clip,keepaspectratio]{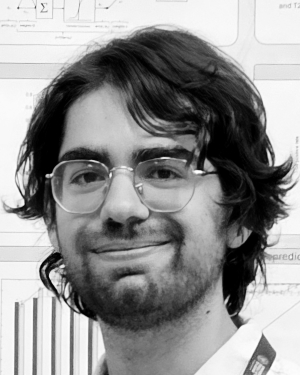}}]{Daniele Lizzio Bosco} was born in Caltagirone, Italy, in 1998. He received his B.S. degree in Mathematics from the University of Udine, Italy, and a double M.S. degree in Artificial Intelligence and Cybersecurity from the University of Udine and the University of Klagenfurt, Austria.
He is currently enrolled in the National PhD program in Artificial Intelligence, a joint PhD program at the University of Udine and the University of Naples Federico II (Italy), and is a member of the Artificial Intelligence Laboratory of Udine (AILAB Udine).
His main research interests include Quantum Machine Learning, Quantum Optimization and Deep Learning applied to Environment and Agriculture. 
He is part of the organization of the annual European Summer School on Quantum AI (EQAI). 
\end{IEEEbiography}

\begin{IEEEbiography}[{\includegraphics[width=1in,height=1.25in,clip,keepaspectratio]{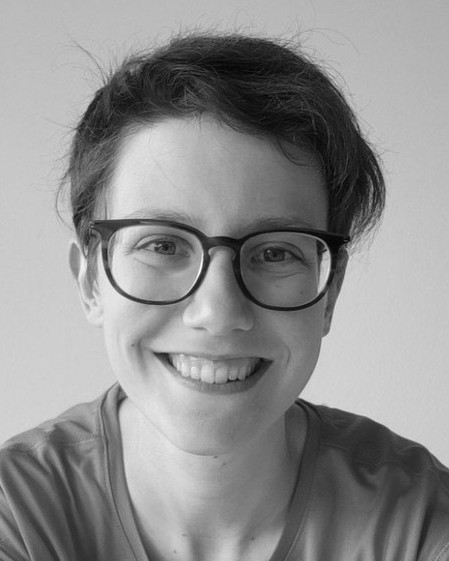}}]
{Beatrice Portelli} is a PhD Student in the National PhD program in Artificial Intelligence, a joint PhD program at the University of Udine and the University of Naples Federico II (Italy), and a member of the Artificial Intelligence Laboratory of Udine (AILAB Udine).
She works employing Deep Learning and Natural Language Processing techniques. She has worked on several projects related to Language Models for Adverse Drug Event extraction and normalization from social media texts, as well as Fact Verification Models. Her current research interests are Machine Learning and Deep Learning methods for risk management in the agricultural-forestry sector.
\end{IEEEbiography}

\begin{IEEEbiography}[{\includegraphics[width=1in,height=1.25in,clip,keepaspectratio]{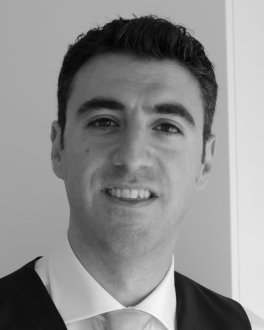}}]{Giuseppe Serra} is an Associate Professor at the University of Udine and he is leading the Artificial Intelligence Laboratory of Udine (AILAB Udine). His research interests include machine learning and deep learning. He was a Lead Organizer of the International Workshop on Egocentric Perception, Interaction and Computing (EPIC), in 2016 and 2017 (ECCV’16—ICCV’17), and he gave tutorials at two international conferences (ICPR’12 and CAIP’13). He also serves as an editorial board for IEEE Transactions on Human Machine Systems and ACM TOMM. He was a technical program committee member of several conferences and workshops.
\end{IEEEbiography}

\EOD

\end{document}